\documentstyle[aps,prb,eqsecnum]{revtex}

\begin{document}
\draft

\title{Effective action for Superconductors and BCS-Bose crossover}

\author{S. De Palo, C. Castellani, C. Di Castro}
\address {Istituto Nazionale di Fisica della Materia e
Dipartimento di Fisica, Universit\'a La Sapienza,} 
\address{Piazzale A. Moro, 00185 Roma, Italy}
\author{B. K. Chakraverty}
\address {Laboratoire d'\'Etudes des Propri\'et\'es Electroniques des
Solides, C.N.R.S.,B.P. 166, 38042 Grenoble Cedex 9, France}

\maketitle

\begin{abstract}

A standard perturbative expansion around the mean-field solution
is used to derive the low-energy effective action for superconductors 
at $T=0$. Taking into account the density 
fluctuations at the outset we get the effective action where the density 
$\rho$ is the conjugated momentum to the phase $\theta$ of the 
order parameter. In the hydrodynamic regime, the dynamics of the superconductor
is described by a time dependent non-linear Schr\"odinger equation (TDNLS) for the field $\Psi(x)=\sqrt{\frac{\rho}{2}}e^{i\theta}$. 
The evolution of the density 
fluctuations in the crossover from weak-coupling (BCS) to strong-coupling 
(Bose condensation of localized pairs) superconductivity is discussed for 
the attractive Hubbard model. In the bosonic limit, the TDNLS equation 
reduces to the the Gross-Pitaevskii equation for the order parameter, as 
in the standard description of superfluidity. 
The conditions under which a phase-only action can be derived in the 
presence of a long-range interaction to
describe the physics of the superconductivity of ``bad metals'' are 
discussed.
\end{abstract}


\section*{Introduction}

The description of a superconducting system at low temperatures by 
an effective action acquired a renewed interest due to the recent 
attention to vortex dynamics \cite{Vort} and to the non-standard 
behavior \cite{Rass} of the High $T_c$ Superconductors (HTCS). 
The advantage in having an effective theory is well known: the complex 
microscopic theory is replaced by a simpler one, written in terms 
of relevant variables only. As an example, near the critical temperature,
the equilibrium \cite{Land} and non-equilibrium \cite{None,Tsun,Klei} 
properties of a superconductor are well described by the Landau-Ginzburg 
theory that is written in terms of the complex order parameter
$\Delta(x)=\langle \phi_{\downarrow}(x)\phi_{\uparrow}(x)\rangle$.

At zero temperature, the effective theory which 
allows a minimal description of the low-energy 
properties of a superconductor (from the Meissner effect 
to the flux quantization) is written in terms of the phase $\theta(x)$ 
of the order parameter. Microscopic derivations of such an 
effective theory are usually carried out by deriving the low temperature
effective action 
in terms of the complex order parameter. This effective action then
leads to a wave-equation for the phase while the order parameter 
itself obeys a non-linear equation with second order time derivatives 
\cite{Tsun,Kemo,Stoo}. 

By contrast, the low-energy dynamics of a bosonic superfluid is well described
by a Gross-Pitaevskii equation \cite{Gros} {\it i.e.} a time-dependent
non-linear Schr\"odinger equation (TDNLS) for the complex order parameter
$\Psi=|\psi| e^{i\theta}$. The continuity equation implied for 
$|\psi(x)|^2$
by the TDNLS description is a continuity 
equation for the full density which, in this case, 
can indeed be identified 
with $|\Psi(x)|^2$. For a fermionic superfluid 
there is instead a non-trivial relation between the density $\rho(x)$ and 
the order parameter $\Delta(x)$. 
However, the Gross-Pitaevskii description is expected to hold
also for fermions in the strong pairing limit, when the Cooper pairs, 
that are strongly overlapping in the BCS description, can be considered as 
composite bosons. In this limit, the phenomenological effective lagrangian 
proposed by Feynman \cite{Feyn}, suggesting a description of the 
superconducting system as a bosonic superfluid, should be recovered.

Recently, Aitchson {\it et al.} \cite{Aitc} and M.Stone \cite{Ston} 
re-interpreted the wave-like equation for the phase of the order parameter for 
a BCS superconductor in terms of a non-linear Schr\"odinger equation.
They reconciled the wave-equation with a TDNLS description by re-writing the 
low-energy effective lagrangian for $\Delta(x)$ in terms of 
the phase and of the field that plays the role of the conjugated momentum of $\theta(x)$ {\it i.e.} the local density $\rho(x)$.
The field $\rho(x)$ is however introduced in the theory 
as an {\it ad hoc} field. This procedure needs indeed an identification 
{\it a posteriori} of the "wave-function" obeying the non-linear Schr\"odinger equation and does not explain the strong coupling limit.

Here we give a {\it microscopic} derivation of the zero-temperature 
low-energy effective action in terms of relevant physical fields, 
including from 
the beginning the density $\rho(x)$ together with the complex order 
parameter $\Delta(x)$ \cite{Dica}. Our derivation is quite general. 
We consider, 
as usual, small fluctuations of the fields from their mean field values 
and retain in the effective action only the leading momentum and 
frequency dependences (hydrodynamic limit).

The effective action is obtained 
by a standard perturbative procedure \cite{Klei} with two main differences 
with respect to previous approaches. First, we decouple the pairing term by 
means of an Hubbard-Stratonovich (HS) transformation in the particle-particle
{\it and} in the particle-hole channels to include both fluctuations. 
Since the auxiliary fields introduced by the
Hubbard-Stratonovich transformation are not the physical fields, 
the latter fields are introduced
by a further transformation. This makes the low-energy effective
lagrangian easily readable and allows for a direct interpretation of the 
local density as the conjugated momentum of the phase. Furthermore
the correct strong pairing limit is straightforward.

The equations of motion supported by this microscopic derivation are
the non-linear Schr\"odinger equation for the variable
$\Psi(x)=\sqrt{\frac{\rho(x)}{2}}e^{i\theta(x)}$, in agreement with
Ref.\cite{Aitc} and \cite{Ston}, and an equation for
the modulus of the order parameter $|\Delta(x)|$ relating the local 
condensate to the local density. 

When we consider the BCS-Bose crossover for the attractive Hubbard model,
the expression of the lagrangian in terms of the physical fields shows 
that, 
on increasing the pairing interaction, 
the density tends to fluctuate coherently
with $|\Delta(x)|^2$. In particular, in the 
low-density limit, the two fields become proportional and the 
description of low-energy dynamics by means of two equations of motion 
becomes redundant. The low-energy dynamics is then described by 
a Gross-Pitaevskii equation (a TDNLS equation for the complex order 
parameter $\Delta(x)$).

On various occasions the crossover between the BCS (weak-coupling) and
the Bose (strong-coupling) limit has been studied \cite{Rand,Rann,Haus}
and the analysis of this problem has acquired particular relevance for the
High T$_c$ Superconductors. Indeed, some common characteristics of these
compounds (the small value of $k_F\xi \simeq 10-20$, compared to the BCS
values, and the small carrier density \cite{Coer}) can be in favor of
a situation close to condensation of localized Cooper pairs, especially
in the under-doped HTCS. In this context, the role of the phase fluctuations
for this compounds has been analyzed looking at the phase-only 
action in the presence of a long-range Coulomb force \cite{Chak,Eme1}.
The low-energy action which we obtain, including the density fluctuations 
together with the fluctuations of the complex order parameter, easily allows
to study situations that can deviate from the BCS behavior. Therefore, at the 
end, we also consider long-range forces in our approach and compare 
(where possible) our result with the Emery and Kivelson proposal 
\cite{Eme1} of a phase-only effective action for a description of 
HTCS as "bad metals".

The paper is organized as follows.
The first section is dedicated to the derivation of the low-energy effective
lagrangian in terms of the physical fields for a simple continuum model.
In section II, we derive the equations of motion and obtain the non-linear 
Schr\"odinger equation for the field 
$\Psi(x)=\sqrt{\frac{\rho(x)}{2}}e^{i\theta(x)}$
In the third section we consider the explicit crossover from weak to strong 
coupling starting from the attractive Hubbard model. Finally, the effects of 
a long-range interaction are specifically considered in the last section 
where the phase-only action is also derived.

\section{The effective action}

The details of the interaction inducing pairing are not
relevant for our purposes. We then consider a very simple model: a fermionic
gas, in $3$-$d$ dimensions, with a local pairing interaction and an energy 
cut-off to remove ultraviolet divergences. The ``partition function''
at $T=0$ 
($\hbar=k_B=1$) for this system is:
$${\cal Z}=\int {\cal D}\phi^{+}(x) {\cal D}\phi(x)
e^{iS(\phi^{+}(x),\phi(x))}$$
\begin{equation}
 S=\sum_{\sigma}\int d^{4}x \phi_{\sigma}^{+}(x)
\left[ i\partial_{t}+
{\bigtriangledown^{2}\over 2m}+\mu 
\right] \phi_{\sigma} (x)
-U \int d^{4}x \phi^{+}_{\uparrow}(x)\phi_{\uparrow}(x)
\phi^{+}_{\downarrow}(x)\phi_{\downarrow}(x)
\label{BCS}
\end{equation}

\noindent where $x=(t,\vec{r})$ ,$\phi_{\sigma}^{+}(x)$ 
and  $\phi_{\sigma}(x)$ are the Grassmann variables associated to fermions of 
spin $\sigma$, $\mu$ is the chemical potential and $U \le 0$ is the attractive 
interaction \cite{Tfin}. 

To obtain the effective action in terms of relevant variables for the 
low-energy region, we decouple the interaction term by means 
of an HS transformation in both the particle-particle 
and the particle-hole channel, introducing in this way the auxiliary fields 
associated to the complex order parameter and to the density. The decoupling 
of the pairing term reads:
$$exp\left[  -iU \int d^{4}x \phi^{+}_{\uparrow}(x)\phi_{\uparrow}(x)
\phi^{+}_{\downarrow}(x)\phi_{\downarrow}(x)\right]\propto
\int {\cal D}\rho^{_{HS}}{\cal D}\rho^{_{HS}}_{spin}
{\cal D}\Delta^{_{HS}}{\cal D}(\Delta^{_{HS}})^{+}exp \left\{ \right.
$$
$$
-i\alpha \int d^4q \int d^4k 
\rho^{_{HS}}(q)  \left[ \phi^{+}_{\uparrow}(k)\phi_{\uparrow}(k-q) +
\phi^{+}_{\downarrow}(k)\phi_{\downarrow}(k-q) \right]+
$$
$$+\alpha \int d^4q \int d^4k \rho^{_{HS}}_{spin}(q)
 \left[ \phi^{+}_{\uparrow}(k)\phi_{\uparrow}(k-q) -
\phi^{+}_{\downarrow}(k)\phi_{\downarrow}(k-q) \right]+
$$
$$-i\gamma \int d^4k \int d^4q \left[ \Delta^{_{HS}}(q)
\phi^{+}_{\uparrow}(k)\phi^{+}_{\downarrow}(-k-q) + 
\left(\Delta^{_{HS}}(q)\right)^{+} \phi_{\downarrow}(k)\phi_{\uparrow}(-k-q) 
\right]+
$$
\begin{equation}
+ i\int d^4q \left.\frac{1}{U} \rho^{_{HS}}(q) \rho^{_{HS}}(-q) 
+\frac{1}{U} \rho^{_{HS}}_{spin}(q) \rho^{_{HS}}_{spin}(-q) 
+\frac{1}{U} (\Delta^{_{HS}}(q))^{+} \Delta^{_{HS}}(-q) \right\}
\label{DISA}
\end{equation}
where the decoupling in the particle-hole channel has been made
following the Hamman prescription: $-U\phi^{+}_{\uparrow}(x)
\phi_{\uparrow}(x)\phi^{+}_{\downarrow}(x)\phi_{\downarrow}(x)=-
\frac{U}{4} 
\left[ \left( \phi^{+}_{\uparrow}(x)\phi_{\uparrow}(x)+ \right.\right.$
$\left. \left.\phi^{+}_{\downarrow}(x)\phi_{\downarrow}(x)\right)^2-\left( 
\phi^{+}_{\uparrow}(x)\phi_{\uparrow}(x)-\phi^{+}_{\downarrow}(x)
\phi_{\downarrow}(x)\right)^2\right]$ \cite{Hamm}.
The relative weights $\alpha$ and $\gamma$ of the p-p and p-h channels are 
arbitrary but for the ``completeness''condition $\alpha^2$+$\gamma^2$=$1$. 
All the resulting effective actions are equivalent when treated exactly 
\cite{Cadi}
allowing that each channel contributes to the other one 
at large $q$. However, to have the 
mean-field results as a starting point, we disregard the ``completeness''
condition and take $\alpha=\gamma=1$ \cite{over} while introducing a cut-off 
in the momentum space to prevent over-counting. 
Indeed, a restricted integration on $q \leq q^* $ separates 
the contributions related to the two channels. Moreover, from now on, we will 
not consider the field $\rho_{spin}(x)$ since, in the low-energy limit, 
spin density fluctuations are completely decoupled from the rest of the 
system. 

In the following we briefly report the main steps of the derivation of the 
effective action, while its full derivation is reported in the Appendix.
As a first step, since the HS fields 
$\rho^{_{HS}}, \Delta^{_{HS}}, (\Delta^{_{HS}})^{+}$ act as
conjugate to the physical fields and coincide with them only at the saddle 
point level \cite{NOT1}, we introduce the physical fields by means of a
further ``Gaussian'' transformation of the quadratic terms appearing in 
(\ref{DISA}).
Secondly, to explicit the dependence on the gradients of the phase $\theta$ 
\cite{Kemo,Aitc} of order parameter in the action (\ref{FEHU}), we make a
gauge transformation $\phi(x)\rightarrow \phi(x) e^{i\theta(x)/2}$.

To obtain the effective action we retain the first and the second order
of a standard expansion  \cite{Klei} of the effective bosonic action 
(\ref{BOSO}) which is obtained integrating out the fermionic fields, around the 
saddle-point solution in the superconducting state.
As usual, we assume that the fluctuations of the fields around their 
saddle point values are small and in particular that the variations of 
the phase in space and time are slow. The last step consists in integrating 
out the HS fields in (\ref{HUST}).

The effective action in terms of the physical fields reads:
$$
S_{eff}(\rho,|\Delta|,\theta)= -\int d^{4}x
\rho(x)\left[ \partial_{t}\theta(x)/2 +\frac{(\bigtriangledown
\theta)^{2}}{8m} \right] +
$$
\begin{equation}
-\int \frac{d^{4}q}{(2\pi)^{4}} \left\{ \delta\rho_q \Pi_{\rho\rho}(q)
 \delta\rho_{-q} +\delta|\Delta|_q\Pi_{\Delta\Delta}(q) \delta|\Delta|_{-q}+
2\delta\rho_q\Pi_{\rho,\Delta}(q)\delta|\Delta|_{-q} \right\}
\label{EFAC}
\end{equation}
where  
$$\Pi_{\rho\rho}(q)=\frac{1}{4} 
\left[ \frac{1}{A_{q}-D_{q}}+U- \frac{[ B_{q}+B_{-q}]^{2}}
{(A_{q}-D_{q})^{2}I_{q}} \right],$$
$$\Pi_{\Delta\Delta}(q)=-\left[ \frac{1}{I_{q}}-U \right],$$
$$\Pi_{\rho \Delta}(q)=\frac{[B_{q}+B_{-q}]}
{2(A_{q}-D_{q})I_{q}}
$$
with
$$I_{q}= \frac{[B_{q}+B_{-q}]^{2}}{A_{q}-D_{q}} -\left[ D_{q}+\frac{1}{2}(
C_{q}+C_{-q}) \right]$$

The explicit expressions of the above coefficients at $T$=$0$ are reported 
in the Appendix. 

It is easily checked that eq.(\ref{EFAC}) reproduces the generalized RPA
approximation for the response function in the low-frequency and
low-momentum limit \cite{Rann}. The low-energy and low-momentum
excitations of a superconducting system described by a microscopic
action like (\ref{BCS}) are density fluctuations: {\it i.e.}
the Anderson-Bogolubov sound mode \cite{Bogo,Ande}. In order to describe
this collective mode, it is sufficient to consider only the hydrodynamic
limit of the effective action (\ref{EFAC}) by taking the
$\vec{q},\omega \rightarrow 0$ limit of the coefficients of
the quadratic part 
in $\rho$ and $|\Delta|$. Notice that at zero temperature the 
static ($\omega=0, \vec{q}\rightarrow 0$) and the dynamic limit 
($ \vec{q}= 0,\omega\rightarrow 0$) of the coefficients in (\ref{EFAC}) 
coincides.

In the BCS limit we recover the usual dispersion for Anderson-Bogolubov mode 
\cite{Bogo,Ande}: 
$\omega^{2}-v_{F}^{2} \left[1+UN_{0}\right]{\bf q}^2/3=0$ where
$v_{F}=k_{F}/m$ is the Fermi velocity and $N_0$ is the density of states per spin at
the Fermi energy. Notice that, in the BCS limit, $\delta\rho(x)$ and 
$\delta|\Delta|(x)$ are decoupled since $\Pi_{\rho\Delta}^0=0$. 
Indeed, $B_0\propto \int d\epsilon N(\epsilon)\epsilon$ vanishes 
due to the particle-hole symmetry property of the BCS (weak-coupling) theory.
Contributions beyond the hydrodynamic limit ({\it i.e.}
a gradient expansion of the coefficients in the action (\ref{EFAC}))
represent higher order corrections to the sound mode dispersion. 

Making a direct comparison with previous derivations \cite{Aitc,Ston}
we find that the density $\rho_0$ in front of the phase 
term has become the full local density $\rho(x)=\delta\rho(x)+\rho_0$, while 
higher order contributions in $\left[ \partial_{t}\theta(x)/2 
+\frac{(\bigtriangledown\theta)^{2}}{8m} \right]$ 
are eliminated away under the integration of the HS fields. 
A direct way to obtain this result is to make an appropriate 
translation of fields in the action (\ref{BOSO}) 
which is expressed in terms of both the HS fields and the
physical fields. We use the following transformation of unitary Jacobian:
\begin{equation}
\tilde{\rho}^{_{HS}}(x)=\rho^{_{HS}}(x)+\left[ \partial_{t} \theta(x)/2 
+(\bigtriangledown \theta(x))^{2}/8m\right]
\label{tra}
\end{equation}
and rewrite the term $\rho^{_{HS}}(x)\rho(x)$ in the action (\ref{BOSO}) as $\rho^{_{HS}}(x)\rho(x)=\left\{
\tilde{\rho}^{_{HS}}(x)-\left[ \partial_{t} \theta(x)/2 +(\bigtriangledown 
\theta(x))^{2}/8m\right] \right\} \rho(x) $. 
The second term of this expression is recognized as the first contribution
in the action (\ref{EFAC}). Then the integration over 
$ \tilde{\rho}^{_{HS}}(x)$ and the other HS fields (within the same 
approximations used previously) results into the phase independent 
contribution of (\ref{EFAC}). By means of this procedure, we see that the first
term of (\ref{EFAC}) is always present in the full action.
This result strongly relies on the introduction of the density fluctuations at
the outset of our approach and on the use of the physical fields
$\rho$ and $\Delta$.

The transformation (\ref{tra}) also allows us to see that
the dependence of the action on $\partial_t\theta (x)/2$ shown
in eq.(\ref{EFAC}) is not restricted to the Gaussian approximation
for the HS fields showing that $\rho(x)$ is the proper coefficient of
$\partial_t\theta (x)/2$. 
This approach has to be contrasted to the one of Ref.(11), where the density 
$\rho(x)$ is derived phenomenologically by reconstructing 
$\rho(x)=\rho_0+\delta\rho(x)$ via the Galilean invariance from mean field
like equations. \cite{NOGA} 

\section{The low-energy effective action: the non-linear
Schr\"odinger equation}

Let us consider, for completeness, the system in 
the presence of an electro-magnetic field $ A\equiv(\varphi(x),\vec{A}(x))$ that we 
introduce in (\ref{EFAC}) through the minimal substitution on the phase
$\partial_{t} \theta/2 \rightarrow \partial_{t}\theta/2+e\varphi$, 
$\bigtriangledown\theta/2 \rightarrow\bigtriangledown\theta/2-
\frac{e}{c}\vec{A}$ 
and adding the free electromagnetic action. The effective action then reads:
$$ 
S_{eff} = -\int d^{4}x
\left\{ \rho(x) \left[ \partial_{t} \theta(x)/2 
+e\varphi(x) +\frac{1}{2m} \left( \frac{e}{c}{\vec A}(x)-
\frac{\bigtriangledown \theta (x)}{2} \right)^{2}  \right]+ \right.
$$
$$ 
+\left[ \Pi_{\rho\rho}^{0}\delta\rho(x)^2 +\Pi_{\Delta\Delta}^{0}
\delta|\Delta(x)|^{2}+2\delta\rho(x)\Pi_{\Delta\rho}^{0}
\delta|\Delta(x)| \right] + 
$$
\begin{equation} 
\left. +\frac{1}{8\pi} \left[ (\bigtriangledown \varphi(x)+
\frac{1}{c} \partial_{t}{\vec A}(x))^{2}+ (\bigtriangledown \wedge 
{\vec A}(x))^{2} \right] \right\}
\label{EFLA}
\end{equation}
where we call 
$\Pi_{\rho\rho}^{0},\Pi_{\rho\Delta}^{0},\Pi_{\Delta\Delta}^{0}$
the $q \rightarrow 0$ limit of the coefficients previously defined.

We have already discussed the origin of the first term in (\ref{EFAC}).
Because of this term the 
role of physical density is directly stated: the functional derivative 
of (\ref{EFLA}) with respect to $\partial_{t}\theta(x)$ gives $-\rho(x)/2$ 
and identify half of the physical density as the conjugated momentum 
to the phase.

This feature suggests to introduce the field
$\Psi(x)=\sqrt \frac{ \rho(x) }{2} e^{i\theta(x)}$ and to re-write an
effective action in terms of it. The eq. (\ref{EFLA}) becomes
$S_{eff}$=$\int d^4x {\cal L}_{eff}$ with:
$${\cal L}_{eff}= \frac{i}{2} \left[ \Psi^{*} 
\partial_{t} \Psi - \Psi \partial_{t} \Psi^{*} \right] 
-2e\varphi \Psi^{*} \Psi
-\frac{1}{4m} \left(i\bigtriangledown-\frac{2e}{c}{\vec A}  \right)
\Psi^{*} \left(- i\bigtriangledown -\frac{2e}{c}{\vec A} \right) \Psi+ $$
$$
-4\Pi^0_{\rho\rho}(\Psi\Psi^{*}-\rho^{0}/2)^{2}
-\Pi^0_{\Delta\Delta}\delta|\Delta|^{2}
-4\Pi^0_{\rho\Delta}(\Psi\Psi^{*}-\rho^{0}/2)\delta|\Delta|
$$
\begin{equation} 
-\frac{1}{8\pi}\left[ (\bigtriangledown \varphi+
\frac{1}{c} \partial_{t}{\vec A})^{2}+ (\bigtriangledown \wedge 
{\vec A})^{2}\right] 
\label{EFFE}
\end{equation}
Notice that the presence of a ionic background, as usually assumed to 
conserve the charge neutrality of system, leads to the replacement in 
(\ref{EFFE}) of $2e\varphi(x) \Psi^{*} \Psi$ in 
$e\varphi(x)\left( 2\Psi^{*} \Psi-\rho_{ion} \right)$.  
This term vanishes on average since $\rho_0=\rho_{ion}$, but it becomes 
$e\varphi(x)\left( 2\Psi^{*} \Psi-\rho_0 \right)$ by retaining the density 
fluctuations.

The mapping of the low-energy effective action (\ref{EFLA}) into
(\ref{EFFE}) holds in the strict hydrodynamic limit. Indeed only the term
$\frac{\rho}{8m}\left(\bigtriangledown \theta \right)^2$
contained in $\frac{1}{4m} \bigtriangledown \Psi^{*} \bigtriangledown \Psi$
is present in the lagrangian (\ref{EFLA}), while the remaining term
$\frac{1}{32m\rho}\left(\bigtriangledown \rho\right)^2 $ should be evaluated
from the gradient expansion of the quadratic contribution $\propto (\delta\rho)^2$
in (\ref{EFAC}). However the coefficient 
of $(\bigtriangledown \rho)^2$ we obtain from (\ref{EFAC}) is not
$\frac{1}{32m\rho}\left(\bigtriangledown \rho\right)^2 $
(for instance, this evaluation at half-filling, for the Hubbard model,
gives the same expression $\frac{1}{9m\rho^0}\frac{1}{16}$ 
we would obtain from expanding 
the compressibility at low $q$ in the normal phase). 
Therefore the mapping is reliable only when this difference 
is irrelevant: {\it i.e.} in the low-frequency and low-momentum limit for the
the density fluctuations.

The introduction of the field $\Psi(x)$ allow us to derive the equation
of motion for $\Psi(x)$ in form of a non-linear Schr\"odinger equation.
By taking the functional derivatives of (\ref{EFFE}) with respect to $\Psi^{*}$, 
$\delta|\Delta|$ and $\vec{A}$ we get:
\begin{equation}
i\partial_{t} \Psi(x) -2e\varphi(x)\Psi(x) -\frac{1}{4m}
\left(\frac{2e}{c}{\vec A}(x)+ i\bigtriangledown \right)^{2} 
\Psi(x) -\frac{4}{\chi}(\Psi(x)\Psi^{*}(x)-\rho^{0}/2)\Psi(x)=0
\label{SCNO}
\end{equation}
\begin{equation}
\delta|\Delta(x)|= -2\frac{\Pi^0_{\rho\Delta}}{\Pi^0_{\Delta\Delta}}
(\Psi(x)\Psi^{*}(x)-\rho^{0}/2) 
\label{STAT}
\end{equation}
\begin{equation}
j(x) =\rho(x)\frac{e}{m}\left(\frac{\bigtriangledown \theta(x) }{2}-
\frac{e}{c}{\vec A}(x)\right) 
\label{CORR}
\end{equation}
where $\chi$ represents the compressibility given by:
$$\chi=\frac{1}{2}\frac{\Pi^0_{\Delta\Delta}}
{\Pi^0_{\rho\rho}\Pi^0_{\Delta\Delta}-(\Pi^0_{\rho\Delta})^2}=2
\frac{\left(A_0-D_0\right)\left[1+U\left(D_0+C_0\right)\right]-4UB_0^2}
{\left[1+U\left(A_0-D_0\right)\right]\left[1+U\left(D_0+C_0\right)\right]
-4U^2B_0^2}$$ and the current $j(x)$ obeys the Maxwell equation:
$\bigtriangledown \wedge \vec{B}=\frac{4\pi}{c}j(x)
+\frac{1}{c}\partial_t\vec{E}$.

Going back to the variables $\rho(x)$ and $\theta(x)$, the imaginary and real 
part of the non-linear Schr\"odinger equation (\ref{SCNO}) are
the continuity equation and the equation that 
links fluctuations in the density to fluctuations in the phase respectively:
\begin{equation} 
\partial_{t}\rho(x)+\bigtriangledown \left[ \frac{\rho(x)}{m}
\left(\frac{\bigtriangledown \theta(x)}{2}
-\frac{e}{c}{\vec A}(x)\right)\right] =0  
\label{CONT}
\end{equation}
\begin{equation} 
\delta\rho(x)=-\chi\left\{ \left[ \partial_{t} \theta(x)/2 -e\varphi(x)
\right]+\frac{1}{2m}\left(\frac{\bigtriangledown \theta(x)}{2}
-\frac{e}{c}{\vec A}(x)\right) ^{2} \right\} 
\label{DENS}
\end{equation}
The continuity equation (\ref{CONT}) is equivalently obtained
by setting to zero the functional derivative of (\ref{EFLA}) with respect to
$\theta(x)$. The appearance of the continuity equation is indeed due to the
fact that the phase is the Goldstone 
boson field whose associated equation of motion is the 
charge conservation, provided the electromagnetic 
response is only given in terms of the phase (as it is the zero 
temperature case).

As already pointed out, the difference (in general) between the modulus squared 
$|\Delta|^{2}$ of the order parameter and the density $\rho$ in a
superconductor prevents to obtain a description of the dynamics uniquely in
terms of a Gross-Pitaevskii 
equation. In fact, the non-linear Schr\"odinger equation
(\ref{SCNO}) is related to the continuity equation and cannot be extended to
describe the condensate $|\Delta|$, but for the bosonic limit which we discuss
in the next section. To obtain the dynamics of the condensate we should have 
made a gradient expansion of the coefficients of the effective action 
(\ref{EFAC}), at least
up to the second derivatives (in time and space) of the modulus $|\Delta|$ 
to define the coherence length for this field.
Being our description limited to the hydrodynamic low-energy regime, the
dynamics of the amplitude of the order parameter is missing. Indeed,
{\it the local}  eq.(\ref{STAT}) should be more properly understood as a 
static relation, valid at $q=0$, 
that links the density fluctuations to the 
order parameter amplitude fluctuations.
In conclusion for {\it almost} constant $|\Delta|$, the relevant dynamics is described 
by the non-linear 
Schr\"odinger equation for $\Psi(x)$. The coherent length we get for $\Psi(x)$
is linked to the compressibility and is then roughly of the order 
of inter-particle spacing. 

\section{The Bosonic limit: the attractive Hubbard model}

We want to analyze the problem of the crossover from standard BCS 
(weak-coupling) superconductivity to strong coupling superconductivity, 
where the Cooper pairs are made of couples of strongly bound fermions,
by extending the previous approach to the attractive 
Hubbard model on a lattice \cite{NotV}. 
Our aim is to understand the interplay between density 
and order parameter while increasing the strength of the pairing 
interaction.

In particular, we want to see how the description of the dynamics of 
a strong coupling superconductor evolves from the BCS limit to the 
Gross-Pitaevskii equation and how the role played by the local density 
is taken by the square of the order parameter amplitude, at least 
in the low-density limit. 

The fermionic action for the Hubbard model is:
\begin{equation}
S(\phi^{+}(x),\phi(x))=\sum_{i,\sigma} 
\phi_{i,\sigma}^{+}\left( i\partial_{t} +\mu \right)  \phi_{i,\sigma} + t
\sum_{\langle ij \rangle} \left( \phi_{i,\sigma}^{+}\phi_{j,\sigma}
+\phi_{j,\sigma}^{+}\phi_{i,\sigma} \right)
-U \sum_{i} \phi^{+}_{i,\uparrow}\phi_{i,\uparrow}
\phi^{+}_{i,\downarrow}\phi_{i,\downarrow}
\label{HUBB}
\end{equation}
where, as in the previous section, $U\le 0$, the sum $\langle ij \rangle$
is over the nearest neighbour pairs and the lattice spacing is $a=1$. 

In deriving the effective action for the Hubbard model on a lattice two cases
are of relevance:

i) the hydrodynamic limit where one can assume small variations $|\theta_i-\theta_j|\simeq a|\bigtriangledown\theta_i|
\ll 1,$ for the phase of the order parameter. In this case we follow the same procedure applied in the continuum
limit, in particular we expand $-iTr ln {\hat{G}}^{-1}$ in (\ref{BOSO}) in powers of 
$1-\cos \frac{1}{2}(\theta_{i}-\theta_{j})\simeq \frac{1}{8}(\theta_{i}-\theta_{j})^2$ to get 
the gradient contribution $(\bigtriangledown \theta_{i})^2$.

ii) away from the hydrodynamic limit to allow $|\theta_{i}-\theta_{j}|\simeq 1$.
In this case the differences in the nearest neighbor site phases are not small and the expansion parameter
$|\theta_{i}-\theta_{j}|$  should be substituted by $t/|U|$. Indeed, it is only in the limit $t/U \ll 1$ that one 
can assign a consistent meaning to the site phase $\theta_i$ \cite{Ran2}.
We shall consider this case in the next section, in connection with the derivation of the phase-only action
(\ref{EFCO}).

Here, to analyze the bosonic limit, we hold on the more standard hydrodynamic 
approach. The derivation is straightforward and the effective low-energy and 
low-momentum action in terms of the physical fields reads:

$$ S_{eff}= - \int dx^4 \left\{ [
\rho(x) \partial_{t} \theta(x)/2 -
\frac{\langle E^{Kin}_0 \rangle }{16} ( \bigtriangledown \theta(x) )^2 
] \right.+$$
\begin{equation}
\left. +\left[ \delta\rho(x)\Pi^0_{\rho\rho}\delta\rho(x)+
\delta|\Delta(x)| \Pi^0_{\Delta\Delta}\delta|\Delta(x)|+2
\delta\rho(x)\Pi^0_{\Delta\rho}\delta|\Delta(x)| \right] \right\}
\label{EFHU}
\end{equation}
where  $\Pi_{\rho\rho}^{0},\Pi_{\rho\Delta}^{0},\Pi_{\Delta\Delta}^{0}$ 
are the lattice version of the previously defined coefficients.
Obviously, apart from differences due to the lack of Galilean invariance, 
the effective lagrangians in the continuum (\ref{EFAC}) 
and in the lattice case (\ref{EFHU}) are similar. Indeed,
as far as the explicit time dependence of action is concerned, we obtain the 
same combination $ \rho(x)\partial_{t}\theta(x)/2$ in both cases. The 
stiffness is here given in terms of the averaged kinetic energy
$\langle E^{Kin}_0 \rangle$, 
consistently with the expression of the diamagnetic contribution for 
the current, which changes from $-\rho_0\frac{e^{2}}{cm}$ into 
$\frac{e^{2}}{2c}\langle E^{Kin}_0 \rangle$.

In the attractive Hubbard model, except at half-filling where particle-hole 
symmetry holds, the coefficient, proportional to $B_0$, that links
fluctuations of the modulus of the order parameter 
to the density fluctuations is finite and its role is essential 
for the description of the system in the strong pairing limit.  

In order to get informations on the asymptotic (large $|U|$) behaviour, 
we diagonalize the part of the lagrangian (\ref{EFHU}) which is quadratic
in the fields $\delta\rho$ and $\delta|\Delta|$. We obtain, in this way, two
eigenvectors:  
$$v_{1}= \frac{|\Pi^0_{\rho\Delta}|}{\left[ (\Pi^0_{\rho\Delta})^{2}
+(\Pi^0_{\rho\rho}-\lambda_{1})^{2} \right]^{1/2}}
\delta\rho+
\frac{\Pi^0_{\rho\rho}-\lambda_{1}}{ \left[ (\Pi^0_{\rho\Delta})^{2}
+(\Pi^0_{\rho\rho}-\lambda_{1})^{2} \right]^{1/2} }
\delta|\Delta|$$
$$v_{2}= \frac{\Pi^0_{\Delta\Delta}-\lambda_{2}}{\left[( \Pi^0_{\rho\Delta})^{2}
+(\Pi^0_{\rho\rho}-\lambda_{1})^{2} \right]^{1/2}}\delta\rho+
 \frac{|\Pi^0_{\rho\Delta}|}{\left[ (\Pi^0_{\rho\Delta})^{2}
 +(\Pi^0_{\rho\rho}-\lambda_{1})^{2} \right]^{1/2}}
\delta|\Delta|$$
with eigenvalues: $$\lambda_{1,2}=\frac{1}{2} 
\left\{ (\Pi^0_{\rho\rho}+\Pi^0_{\Delta\Delta}) \mp 
\left[ (\Pi^0_{\rho\rho}-\Pi^0_{\Delta\Delta})^{2}
+4(\Pi^0_{\rho\Delta})^{2} \right]^{1/2} \right\}.$$

On increasing the interaction, one of the eigenvalues ($\lambda_{2 }$) diverges
like $U^{2}$, while the value of the other ($\lambda_1$) stays finite.
This behaviour implies that the eigenvector corresponding to the diverging
eigenvalue, $v_2$, has vanishing fluctuations $\langle v_2\:v_2\rangle \propto
\lambda_2^{-1}$. Therefore
$\delta \rho$ tends to fluctuate coherently with $\delta|\Delta|$
through the coefficients of the linear combination appearing in $v_{2}$:
\begin{equation}
\delta|\Delta| \simeq -\frac{\Pi_{\Delta\Delta}
-\lambda_{2}}{|\Pi_{\rho\Delta}|} \delta\rho. 
\label{PRCO}
\end{equation}
with the ratio $\frac{\Pi_{\Delta\Delta}
-\lambda_{2}}{|\Pi_{\rho\Delta}|}$ staying finite for large $U$, since 
both $\Pi_{\Delta\Delta}$ and $\Pi_{\rho\Delta}$ diverge like
$U^{2}$.
The density and the modulus of the order parameter experience the same
fluctuations and we expect that they play the same role, so that
a description of the low-energy dynamics of system by
means of two equations of motion for $\delta\rho$ and $\delta|\Delta|$
is becoming redundant. To clarify further the evolution of fluctuations of 
$|\Delta|$ we integrate out the density fluctuations and replace 
$2|\Delta_0|\delta|\Delta| \simeq \delta|\Delta|^2$. The resulting  
effective action is:
$$
S(\theta,\delta|\Delta|^2)= 
\int dx^4 \frac{\langle E^{Kin}_0\rangle }{16} (\bigtriangledown \theta(x))^2
-\int dx^4\left\{ \frac{\rho_0}{2}\partial_t \theta(x)\right.+$$
\begin{equation} \left.
-\frac{1}{4\Delta_0} \frac{\Pi_{\rho\Delta}^0}{\Pi^0_{\rho\rho}}
(\partial_{t}\theta(x_{i}))\delta|\Delta(x_{i})|^2
+\frac{\delta|\Delta(x_{i})|^{2}}{2\Delta_0}
\left[ \Pi^0_{\Delta\Delta}-\frac{\Pi^0_{\Delta\rho}\Pi^0_{\rho\Delta}}
{\Pi^0_{\rho\rho}} \right] \frac{\delta|\Delta(x)|^{2}}{2\Delta_0}
-\frac{1}{\Pi^0_{\rho\rho}}\left(\partial_{t}\theta( x)/4
\right)^{2} \right\}
\label{gp}
\end{equation}
On increasing the interaction, independently from the filling, the term
proportional to $\partial_{t}\theta(x)\delta|\Delta(x)|^{2}$
becomes dominant with respect to
$\frac{1}{\Pi^0_{\rho\rho}}\left(\partial_{t}\theta( x)/4\right)^{2}$ which
only gives negligible corrections to the dispersion of the collective mode. 
Moreover, in the low-density limit, the coefficient
$-\frac{\Pi^0_{\rho\Delta}}{4|\Delta_0|\Pi^0_{\rho\rho}}$
reaches the unity on increasing the interaction. This behaviour and the fact
that the mean field value of $\Delta_{0}^{2}$ is $\rho_{0}/2$ \cite{Nozi},
allows us to recognize in the effective action (\ref{gp}) the
contribution $-|\Delta(x)|^2\partial_t\theta(x)$ where
$|\Delta(x)|^2=\rho_0/2+\delta|\Delta(x)|^2$.
In these limits (strong coupling and low-density), the role of the conjugated 
momentum to the phase
is taken by the square of the modulus of the order parameter: indeed,
$|\Delta(x)|^2$ coincides with $\rho(x)/2$.
Using the same reasoning of the continuum limit
we can then get, as a unique equation of motion,
a non-linear Schr\"odinger equation for the field
$\Psi=\sqrt{\frac{\rho}{2}} e^{i\theta}$. Since,
in this case, $\Psi$ coincides with the order parameter $|\Delta|e^{i\theta}$, 
the TDNLS results into the Gross-Pitaevskii equation.

In the weak-coupling limit, the delocalized structure of the Cooper pairs 
does not permit such a description in terms 
of a Gross-Pitaevskii equation as in 
a bosonic system. Indeed, we need different equations to govern the dynamics 
of the density and the modulus of the order parameter.
On the contrary, when the pairing interaction is strong enough to form tightly bound Cooper 
pairs and the density is low enough that there is no overlapping between 
them, the difference between the square of the modulus of the fermionic order 
parameter and half of the density becomes negligible and they can be identified:
$|\Delta|^2=\rho/2$. 

\section{ The phase-only action in the presence of a long-range interaction}

We devote this last section to the introduction of a long-range
Coulomb interaction $V_C(q)=4\pi e^2/\bf{q}^2$ in
the system. In the discussion of the 
hydrodynamic action in the continuum case we have already implicitly 
considered a charged system when we introduced the
electromagnetic field. This time we pay explicit attention to the effects 
of the Coulomb interaction on the structure of the resulting effective 
action when expressed in terms of the phase only.

It has been proposed on 
several occasions \cite{Chak,Eme1} that a low value of 
the superfluid density and consequently 
of the phase stiffness, enhances the role of the phase fluctuations. 
It is therefore relevant to study the interplay between the phase 
fluctuations and the low-lying energy modes once the amplitude is kept
constant.

To approach this problem let 
us consider the electron action (\ref{HUBB}) in the presence of the Coulomb 
interaction:
$-\frac{1}{2} \sum_{\sigma,\sigma'}\sum_{i,j} 
V_{C}(\vec{r_i}-\vec{r}_j) \phi_{\sigma}^{+}(x_i)
\phi_{\sigma}(x_i)\phi_{\sigma'}^{+}(x_j)\phi_{\sigma'}(x_j).$ 
We decouple the long-range interaction in the 
Hartree channel and then follow the same procedure as before. 
Finally we obtain the effective action:
$$
S_{eff}(\rho,|\Delta|,\theta)= -\int dt 
\sum_{i}\rho(x_i)\partial_{t}\theta(x_i)/2 +\int dt
\sum_{\langle ij \rangle }
\langle E_{0}^{Kin}\rangle /8 \left(1-\cos({\theta_{i}-\theta_{j})} \right)  +
$$
\begin{equation}
-\int \frac{d\omega}{2\pi}\sum_q \left\{
\delta\rho_q\left[\Pi_{\rho\rho}(q) +\frac{1}{2}V_C(q)\right] \delta\rho_{-q} +
\delta|\Delta|_q\Pi_{\Delta\Delta}(q)\delta|\Delta|_{-q}+
2\delta\rho_q\Pi_{\rho\Delta}(q)\delta|\Delta|_{-q} \right\}
\label{EFCO}
\end{equation}
Notice that the coefficient of the quadratic part in the density fluctuations
has been modified by the presence of the Coulomb interaction according to
RPA prescription. As discussed in Section III, to compare with Ref.(19) and Ref.(33) we consider 
here the discrete form  $ -\sum_{\langle ij \rangle
}\left(1-\cos({\theta_{i}-\theta_{j})} \right)$ of the gradient term.
We recall again that is only in the limit $t/|U|\ll 1$ that we can 
consistently derive this term.
In particular, to get a cosine term which is local in time, we need that 
the characteristic time over which 
$\theta_i(t)$ varies is much larger than the inverse quasiparticle gap.


By integrating out all the fields but the phase in the action (\ref{EFCO}), 
we get the phase-only action:
\begin{equation}
S(\theta)=\frac{1}{8}\int \frac{d\omega}{2\pi}\sum_{q}
\:\frac{\Pi(q)}{1+
V_C(q)\Pi(q)}
\;\omega^{2}\theta_q\theta_{-q}+\int dt\sum_{\langle ij \rangle}
\langle E_{0}^{Kin}\rangle/8
\left( 1-\cos(\theta_{i}-\theta_{j}) \right)
\label{EFPH}
\end{equation}
where $$\Pi(q)= \frac{1}{2}
\frac{\Pi_{\Delta\Delta}}
{\Pi_{\rho\rho}\Pi_{\Delta\Delta}-\Pi^2_{\rho\Delta}}.$$
$\Pi(q)$ reduces to the ``short-range'' compressibility $\chi$ in the
limit $q \rightarrow 0$ .

In agreement with Ref.(33), the action is made of two contributions:
the quantistic one ($\propto \omega^2\theta_q\theta_{-q}$), whose coefficient
is proportional, in the static limit, 
to the {\it full} compressibility in the presence of the Coulomb interaction 
(and it is therefore vanishing as $\bf{q}^2$),
and the classic term ($\propto cos(\theta_{i}-\theta_{j})$) 
whose coefficient is the stiffness related to the
superfluid density ($\rho_s/m^*$).

The above phase-only effective action is obtained for a model with long-range
interaction and local pairing interaction without any {\it a priori}
assumption. At $V_C=0$, the action is equivalent to a
$x-y$ quantum model in $d+1$ dimensions. On increasing the pairing
interaction the stiffness decreases due to the enhancement of the effective
mass, while the compressibility increases since we are reaching the bosonic 
limit, where the chemical potential does not depend on the density \cite{Nozi}.

For $V_C \neq 0$, the low-momentum and 
frequency behaviour of the quantistic term is completely dominated by the 
divergence of the long-range interaction  and the vanishing of the 
compressibility. In this regime the action becomes:
\begin{equation}
S\simeq \int \frac{d\omega}{2\pi}\sum_{q}q^{2}\omega^{2} \frac{1}{32\pi e^{2}}
\theta(q)\theta(-q) +\int dt \sum_{\langle ij \rangle }
\langle E_0^{Kin}\rangle/8 \left(1-\cos(\theta_{i}-\theta_{j}) \right)
\label{latt}
\end{equation}
In this case, the sound mode 
corresponding to Anderson-Bogolubov \cite{Bogo,Ande} mode becomes massive
with the plasmon frequency given by $\omega_p^2=4\pi e^2 \rho_0/m^{*}$ in the
continuum and by $\omega_p^2=-2\pi e^2 \langle E_0^{kin}\rangle$ in the
lattice model. 

We can extend the analysis to a superconducting system which is coupled
to a normal metal through the Coulomb interaction.
The density fluctuations of the metal will screen the Coulomb 
interaction in the Hartree channel and affect the coefficient of the 
quantistic part. In this case, the coefficient
of quantistic part of the effective phase-only action is:
\begin{equation}
\Pi(q)/\left[1+ \frac{V_C(q)}{\epsilon(q)}\Pi(q)\right] 
\label{CO2F}
\end{equation}
where $\epsilon(\omega,\vec{q})=1+V_C(q)\Pi_{m}(q)$ and $\Pi_{m}(q)$ is 
the polarization bubble in the metal. This amounts 
to replace the Coulomb interaction
(in the superconductor) with the effective RPA Coulomb
interaction screened by the metal. The low-energy and momentum
limit of (\ref{CO2F}) is $\frac{1}{4e^2\pi}\epsilon(\omega,\vec{q}){\bf q}^2$. 
In the ``static'' region ($\omega \ll |q|$)
the dielectric function is proportional to $1/\bf{q}^2$, the
quantistic coefficient (\ref{CO2F}) is a constant and does not modify
the behaviour of the short-range ($V_C=0$) action (\ref{EFPH}). 
On the contrary, in the dynamic 
limit the dielectric constant is given by $\epsilon \Rightarrow
1+4\pi i\frac{\sigma(\omega)}{\omega}$ and the quantistic coefficient depends on 
the $\omega$ behaviour of the conductivity $\sigma(\omega)$. 

Assuming that an action like this also holds at finite temperature 
(with the same expressions for the coefficients) one recovers the
Emery and Kivelson proposal \cite{Eme1} for a phase-only action describing
HTCS as ``bad metals''. In this case the onset of the locking of the 
coherent superconductive phase would be affected by the $\omega$-behaviour 
of the conductivity.
It is tempting to suggest that, near $T_c$, the role of the external screening 
system can also be played by the ``normal'' component of the  superconducting 
system at finite temperature.

\section{Concluding Remarks}

The procedure we have followed to derive the effective (hydrodynamic) 
action for a superconducting system at $T=0$ strongly relies on 
the introduction of 
the physical fields $\rho,\Delta,\Delta^+$ beside the auxiliary fields 
$\rho^{_{HS}},\Delta^{_{HS}},(\Delta^{_{HS}})^+$ introduced to decouple 
the interaction by means of the Hubbard-Stratonovich transformation. 
The resulting effective action is a proper starting point 
for selecting relevant features in systems that deviates from the standard BCS 
superconductors.

The introduction from the outset of the physical density 
makes it transparent in the action that $\rho(x)$ plays the role of the field
conjugated momentum to the phase. As a direct consequence
the equation of motion for the field $\Psi(x)=\sqrt{\rho(x)/2}e^{i\theta(x)}$
has the form of a non-linear Schr\"odinger equation.
However, besides this equation, in general, also 
the equation of motion for the order parameter amplitude must be considered. 

Our procedure allows to follow the evolution of the
density fluctuations as the pairing interaction is increased, towards the
strong-coupling limit when the Cooper pairs can be considered as
true composite bosons. In this case the density fluctuations are identified
with the square amplitude fluctuations and in the low-density limit
the two fields $\rho(x)/2$ and $|\Delta(x)|^2$ coincide. As a consequence
there is no need of two different equations of motion for $\rho(x)$ and 
$|\Delta(x)|$ as in the BCS case, and the
low-energy behaviour can be described by a Gross-Pitaevskii equation for
the field $\Psi(x)$ which in this case coincides with the order parameter
$\Delta(x)$.

The inclusion of the long-range Coulomb interaction is 
straightforward and the modifications of the action follows the RPA 
prescription. Integrating out all the fields but the phase we get the 
phase-only action in agreement with Ref.(33). In this case
a comparison with the proposal of Emery and Kivelson \cite{Eme1} indicates
that their
effective phase-only action with the coefficient of the quantistic term
proportional to the normal conductivity can be obtained if the superconductor 
is coupled to a metallic component.
\vskip0.5cm

This work has been supported by INFM "Progetto di ricerca avanzata" 1996.

\appendix
\section{The effective action} 

Here we report the derivation of the effective action (\ref{EFAC} )
in terms of the physical fields $\rho(x)$,$|\Delta(x)|$ and $\theta(x)$,
following the steps anticipated in the text.

First, after having decoupled the interaction with the HS fields 
$\rho^{_{HS}}$ and $\Delta^{_{HS}}$ we introduce the physical fields 
$\rho$ and $\Delta$ in the action by two Gaussian transformations of the quadratic
terms $(\rho^{_{HS}})^2$ and $(\Delta^{_{HS}})^2$ appearing in (\ref{DISA}).
In the Nambu formalism, the resulting fermionic-bosonic action, containing
both the HS and the physical fields, reads:
$$ S(\eta,\eta^{+},\rho^{_{HS}},\rho,\Delta^{_{HS}},(\Delta^{_{HS}})^{+}
\Delta,\Delta^{+}) = $$
$$ \int dx^{4} \left\{ \eta^{+}(x)
\left[ i\partial_{t} \sigma_{0}\right]\eta(x) 
+\eta^{+}(x) \left[{\bigtriangledown^{2}\over 2m}+\mu-
\rho^{_{HS}}(x) \right] \sigma_{3}\eta(x)+\rho(x)\rho^{_{HS}}(x)+\right.$$
$$
-\eta^{+}(x){\cal R}e\Delta^{_{HS}}(x)\sigma_{1}\eta(x)+
\eta^{+}(x){\cal I}m\Delta^{_{HS}}(x)\sigma_{2}\eta(x) +
$$
\begin{equation} \left.
+\Delta(x)\left(\Delta^{_{HS}}(x)\right)^{+}+\Delta^{+}(x)\Delta^{_{HS}}(x)
-{U\over 4}\rho^{2}(x)
-U|\Delta(x)|^{2} \right\}
\label{FEHU}
\end{equation}
\noindent where $\eta^{+}=\left( \phi^{+}_{\uparrow},
\phi_{\downarrow}\right)$ is the Nambu spinor, $\sigma_{0}$ is 
the unit matrix  and $\sigma_{i}$ is the Pauli matrix vector.
Notice that, integrating out $\rho^{_{HS}}(x)$ we get the constraint
$\delta(\rho(x)-\eta^{+}(x)\sigma_{3}\eta(x))$ that specifies the field
$\rho(x)$ as the local density. Analogously, acting on $\Delta^{_{HS}}$, 
we obtain 
$\delta\left({\cal R}e\Delta-\frac{1}{2}(\phi_{\downarrow}\phi_{\uparrow}
+ \phi_{\uparrow}^{+}\phi_{\downarrow}^{+}\right)$ 
and $\delta\left({\cal I}m\Delta+\frac{i}{2}
(\phi_{\downarrow}\phi_{\uparrow}-
\phi^{+}_{\uparrow}\phi^{+}_{\downarrow})\right)$. 

Next, we get the action directly in terms of the
modulus of the physical order parameter and its phase. To this aim, after 
writing  
$\Delta=|\Delta|e^{i\theta(x)}$ we make the gauge unitary transformation 
$\phi(x)\rightarrow \phi(x) e^{i\theta(x)/2}$ in the action (\ref{FEHU}). 
Then, we remove the dependencies from the phase in the off-diagonal terms and 
in $\Delta(x)^{_{HS}}|\Delta(x)|e^{-i\theta(x)}
+|\Delta(x)|e^{i\theta(x)}\left(\Delta^{_{HS}}(x)\right)^{+}$ 
by defining the new auxiliary field:
$ {\tilde \Delta}^{_{HS}}(x)= \Delta^{_{HS}}(x)e^{-i\theta(x)}.$ 
Such procedure is convenient in the low-energy limit, since it
explicits the dependence on the gradients of $\theta(x)$.

After these manipulations the action reads:
$$ S(\eta,\eta^{+},\rho^{_{HS}},\rho,|\Delta|,\theta,
{\tilde\Delta}^{_{HS}},({\tilde\Delta}^{_{HS}})^{+})=
\int dx^{4} \left\{ \eta^{+}(x)
\left[ i\partial_{t}+i{\bigtriangledown\theta \over 2m}\bigtriangledown \right]
\sigma_{0}\eta(x)+\right.$$
$$ + \eta^{+}(x) \left[
{\bigtriangledown^{2}\over 2m}+\mu-\left( \partial_{t}\theta(x)/2 +
(\bigtriangledown \theta)^{2}/8m \right) - \rho^{_{HS}}(x)
\right]\sigma_{3}\eta(x)+\rho(x)\rho^{_{HS}}(x)+$$
$$+ i\eta^{+}(x)\frac{\bigtriangledown^{2}\theta}{4m}\sigma_0\eta(x)
+\eta^{+}(x){\cal I}m{\tilde\Delta}^{_{HS}}(x)\sigma_{2}\eta(x)+$$
\begin{equation}
\left.
-\eta^{+}(x){\cal R}e{\tilde\Delta}^{_{HS}}(x)\sigma_{1}\eta(x) +
 2|\Delta(x)|{\cal R}e{\tilde\Delta}^{_{HS}}(x)
-{U\over 4}\rho^{2}(x) -U|\Delta(x)|^{2} \right\}
\label{FEBO}
\end{equation}

By integrating out the fermionic fields we obtain  
the following bosonic action:
$$ S(\theta,\rho^{_{HS}},\rho,|\Delta|,{\tilde \Delta}^{_{HS}},({\tilde
\Delta}^{_{HS}})^+)=-iTr ln {\hat{G}}^{-1}$$
\begin{equation}
+ \int dx^{4} \left[ \rho(x)\rho^{_{HS}}(x)
-{U\over 4}\rho^{2}(x)+ 2|\Delta(x)|{\cal R}e{\tilde \Delta}^{_{HS}}(x)-
U|\Delta(x)|^{2}
 \right]
\label{BOSO}
\end{equation}

Here $\hat{G}$ is the matrix operator representing the one-particle
propagator in the presence of the external bosonic fields. 

$$
{\hat{G}}^{-1}=
+\left\{{\bigtriangledown^{2}\over 2m}+\mu- \rho^{_{HS}}-
\left[ \partial_{t}\theta/2 + \frac{(\bigtriangledown \theta)^{2}}{8m}\right]
 \right\} \sigma_{3}+
$$
\begin{equation}
 \left\{ \left[
i\partial_{t} + i\frac{\bigtriangledown \theta}{2m}\bigtriangledown \right]+
i \frac{\bigtriangledown^{2}\theta}{4m} \right\}\sigma_{0}
-{\cal R}e{\tilde \Delta}^{_{HS}}(x)\sigma_{1}-
{\cal I}m{\tilde \Delta}^{_{HS}}(x)\sigma_{2}
\label{prop}
\end{equation}

We re-write $\hat{G}^{-1}=\hat{G}_{0}^{-1} ({1}-\hat{G}_{0}{\Sigma})$
by introducing the saddle-point one-particle propagator $\hat{G}_{0}$ and the
self-energy matrix ${\Sigma}$. One has  
: $${\hat{G}}_{0}^{-1}= i\partial_{t}\sigma_{0}+
({\bigtriangledown^{2}\over 2m}+\mu - \rho_{0}^{_{HS}})\sigma_{3}-
\Delta_{0}^{_{HS}}\sigma_{1},$$
which, in the Fourier space, leads to 
$$ {{\hat G}_{0}}= \left( \begin{array}{cc}G_{0}&F_{0}\\ F_{0}&-G^{-}_{0}
\end{array}\right)=\frac{1}{\omega^2-E_k^2}
\left( \begin{array}{cc}\omega+\epsilon_k & \Delta_0 
\\ \Delta_0 & \omega-\epsilon_k \end{array}\right) $$
with 
$G_0(\omega,\vec{k})$=$G_0^{-}(-\omega,-\vec{k})$ and 
$F_0(\omega,\vec{k})$=$F_0^{*}(\omega,\vec{k})$ .
Here $\Delta_{0}^{_{HS}}=U|\Delta_{0}|$, $\rho_{0}^{_{HS}}=U\rho_0/2$
are the parameters that extremize the action
(\ref{BOSO}). The ``self-energy'' matrix is
given by:
$$ {\Sigma}(x)=\left\{
 \delta\rho^{_{HS}}(x)+
 \left[ \partial_{t}\theta(x)+ {(\bigtriangledown \theta)^{2}\over 8m} \right]
\right\} \sigma_{3}-\left[ i {\bigtriangledown^{2}\theta(x)\over 4m} 
+ i \frac{\bigtriangledown \theta(x)}{2m}\bigtriangledown \right]
\sigma_{0} + $$
\begin{equation}+\delta{\cal R}e{\tilde \Delta}^{_{HS}}(x) \sigma_{1}-
\delta {\cal I}m{\tilde \Delta}^{_{HS}}(x) \sigma_{2}
\label{Sig}
\end{equation}
We take ${\Sigma}$ as an expansion parameter, by assuming, as usual,
small fluctuations of the fields around their saddle point values and small 
variations of the phase in space and time, since the dependency of the action
in $\theta$ is only through its gradients.
We then rewrite $Tr\ln {\hat{G}}^{-1}$ in (\ref{BOSO}) as $Tr 
\ln {\hat{G}}_{0}^{-1}-Tr{{\Sigma}}_n\frac{1}{n}({\hat{G}}_0{{\Sigma}})^{n}$ 
where the factor $({\hat{G}}_{0}{{\Sigma}})^{n}$ is a product of propagators 
connected into a closed loop by the trace, with $n$ insertions of the
fluctuating fields. 

To derive the low-energy effective action we retain 
in (\ref{BOSO}) the first and the second order of the expansion 
in $(\hat{G}_{0}{\Sigma})$ around the superconducting saddle-point: 
$$
S_{eff}(\delta\rho,\delta\rho^{_{HS}},\delta|\Delta|,\theta,
\delta{\cal R}e{\tilde \Delta}^{_{HS}})= -\int dx^{4} \rho_{0}
\left[ \partial_{t}\theta(x)/2 +\frac{(\bigtriangledown\theta)^{2}}
{8m} \right] +
$$
$$
+\int  \frac{dq^{4}}{(2\pi)^{4}} \left\{
-{U\over 4}
\delta\rho_q \delta\rho_{-q} +\delta\rho^{_{HS}}_{-q}\delta\rho_{q} -
 U\delta|\Delta|_q\delta|\Delta|_{-q}+
  \delta|\Delta|_q\delta{\cal R}e{\tilde \Delta}^{_{HS}}_{-q}
+\right.
$$
$$
+\left[ \delta\rho^{_{HS}}_q +\left(-\frac{i\omega}{2}\theta_q+\frac{1}{8m}
\int  \frac{dq^{4}_{1}}{(2\pi)^{4}} \vec{q}_{1}(\vec{q}_{1}+\vec{q})
\theta_{-q_{1}}\theta_{-q_{1}-q} \right) \right] 
$$
$$ \cdot
\left[ A(q)-D(q) \right]
\left[ \delta\rho^{_{HS}}_{-q} +\left(\frac{i\omega}{2}\theta_{-q}
+\frac{1}{8m}
\int  \frac{d^{4}q_{1}}{(2\pi)^{4}} \vec{q}_{1}(\vec{q}_{1}-\vec{q})
\theta_{q_{1}}\theta
_{-q_{1}+q} \right) \right]+
$$
$$
+\left[ \delta\rho^{_{HS}}_q +\left(-\frac{i\omega}{2}\theta_q+\frac{1}{8m}
\int  \frac{d^{4}q_{1}}{(2\pi)^{4}} \vec{q}_{1}(\vec{q}_{1}+\vec{q})
\theta_{q_{1}}\theta_{-q_{1}-q} \right) \right]
\left[ B_q+B_{-q} \right] \delta{\cal R}e{\tilde \Delta}^{_{HS}}_{-q}
$$
$$-2i\left[ \delta\rho^{_{HS}}_q +\left(-\frac{i\omega}{2}\theta_q+\frac{1}{8m}
\int  \frac{d^{4}q_{1}}{(2\pi)^{4}} \vec{q}_{1}(\vec{q}_{1}+\vec{q})
\theta_{q_{1}}\theta_{-q_{1}-q} \right) \right]
\left[ B_q-B_{-q}\right]\delta{\cal I}m{\tilde \Delta}^{_{HS}}_{-q}
$$
$$
+i\delta{\cal I}m{\tilde \Delta}^{_{HS}}_{q}\left[ C_q-C_{-q}\right] 
\delta{\cal R}e{\tilde \Delta}^{_{HS}}_{-q}-
\delta{\cal I}m{\tilde \Delta}^{_{HS}}_{q}
\left[ D_q-\frac{1}{2}(C_q+C_{-q}) \right]
\delta{\cal I}m{\tilde \Delta}^{_{HS}}_{-q}
$$
\begin{equation} +              \left.
\delta{\cal R}e{\tilde \Delta}^{_{HS}}_{q}
\left[ D_q+\frac{1}{2}(C_q+C_{-q}) \right]
\delta{\cal R}e{\tilde \Delta}^{_{HS}}_{-q}      \right\}
\label{HUST}
\end{equation}
where the coefficients \cite{Legg} are:
$$ A(q)= \frac{i}{2} \int \frac{dk^{4}}{(2\pi)^{4}} \left[
G_0(k+q)G_0(k)+ G^{-}_0(k+q)G^{-}_0(k)\right] $$
$$=
-\frac{1}{2}  \int \frac{dk^3}{(2\pi)^3} 
\frac{\epsilon_{+}\epsilon_{-}-E_{+}E_{-}}
{E_{+}E_{-}} \frac{E_{+}+E_{-}}
{(E_{+}+E_{-})^{2}-\omega^{2}}
$$
$$ B(q)=
\frac{i}{2} \int \frac{ dk^{4} }{ (2\pi)^{4} } 
\left[ F_0(k+q)G_0(k)+ G_0^{-}(k+q)F_0(k)\right]
$$
$$=
-\frac{\Delta_{0}}{4} \int \frac{dk^3}{(2\pi)^3}
\left[ \frac{\epsilon_{+}+\epsilon_{-}}
{E_{+}E_{-}}
+ \frac{\omega}{E_{+}E_{-}} \right]
\frac{E_{+}+E_{-}}{(E_{+}+E_{-})^{2}-\omega^{2}} 
$$
$$ C(q)=
-i\int \frac{dk^{4}}{(2\pi)^{4}} G_0(k+q)G_0^{-}(k)
$$
$$
=\frac{1}{2} \int \frac{dk^3}{(2\pi)^3}
\left[ \frac{E_{+}E_{-}+\epsilon_{+}\epsilon_{-}}
{E_{+}E_{-}}(E_{+}+E_{-}) +\omega \left(\frac{\epsilon_{+}}{E_{+}}
+\frac{\epsilon_{-}}{E_{-}} \right) \right] 
\frac{1}{(E_{+}+E_{-})^{2}-\omega^{2}} 
$$
$$D(q)=i\int \frac{dk^{4}}{(2\pi)^{4}} F_0(k+q)F_0(k)
$$
$$=
-\frac{\Delta_{0}^{2} }{2}  \int \frac{dk^3}{(2\pi)^3}
\frac{1}{E_{+}E_{-}} \frac{E_{+}+E_{-}}{(E_{+}+E_{-})^{2}-\omega^{2} } 
$$
and, in our notation:
$$E_{\pm}= \left[  \epsilon_{\pm}^2+|\Delta_0^{_{HS}}|^2  \right]^{1/2} $$
$$ \epsilon_{+}= (k+q)^2/2m-\mu-\rho_0^{_{HS}} \mbox{   and  } 
\epsilon_{-}= k^2/2m-\mu-\rho_0^{_{HS}} $$

Notice that in the $q \rightarrow 0$ limit one gets
$ C_0+D_0= -\frac{1}{U}-(A_0-D_0). $
In the BCS limit one recovers:
$A_q-D_q \simeq \frac{m k_F}{2\pi^2}=N_0$ and $B_0=0$;
while in the strong coupling limit one obtains
$A_0-D_0 \simeq -\frac{4 \Delta_0^2}{U}$ and $B_0=-(1-\rho_0)\Delta_0
\frac{1}{U}$.

Since we are interested in the low-energy regime of the system in the
derivation of (\ref{HUST}) we have neglected 
contributions that gives corrections that go beyond this regime.
We have indeed neglected terms associated 
to the expansion of $-\left[ i {\bigtriangledown^{2}\theta(x)\over 4m}
+i\frac{\bigtriangledown \theta(x)}{2m}\bigtriangledown \right]\sigma_{0}$
whose contribution is proportional to 
$\int d^{4}q \frac{\vec{q}}{2m}\theta_q 
\int d^{4}k \left[(\vec{k}+\vec{q})G_0(k+q)\vec{k}G_0(k) +
(\vec{k}+\vec{q})F_0(k+q)\vec{k}F_0(k) \right]
\frac{\vec{q}}{2m}\theta_{-q}$.
The coefficient of this term
($ \int d^{4}k \left[(\vec{k}+\vec{q})G_0(k+q)\vec{k}G_0(k) +
(\vec{k}+\vec{q})F_0(k+q)\vec{k}F_0(k) \right]$)
represents the paramagnetic current and it
vanishes in the static limit at $T=0$, while, at finite temperature,
it gives the contribution that, due to the presence of the normal component
of the superfluid, reduces the stiffness. At zero temperature the first 
contribution to the action coming from this term is of order 
$(\bigtriangledown^2\theta)^2$ and then negligible in the low-energy limit.
This also shows that the expansion of 
$-\left[ i {\bigtriangledown^{2}\theta(x)\over 4m}
+i\frac{\bigtriangledown \theta(x)}{2m}\bigtriangledown \right]\sigma_{0}$
does not generate time derivatives of $\theta$. This expansion can only give
time derivatives of $(\bigtriangledown^2\theta)$, the first non vanishing
term being $(\bigtriangledown^2 \dot{\theta} )^{2}$.

The last step consists in integrating out the HS fields 
to get the effective action (\ref{EFAC}). 
Since only the first and the second order terms in the expansion in powers of 
$(\hat{G}_{0}{\Sigma})$ were retained, the action (\ref{BOSO}) is Gaussian in
the auxiliary fields even though it contains higher order corrections in the
phase fields. We can therefore straightforwardly integrate the HS fields
to get the action in terms of the physical fields reported in (\ref{EFAC}). 

Notice that 
in this integration we can ignore contributions coming from the coupling of 
$\delta{\cal I}m{\tilde \Delta}^{_{HS}}_{q}$ with the other
HS fields. The corrections coming from this coupling 
to the coefficients of the effective action (\ref{EFAC}) represent 
higher order corrections to the hydrodynamic limit.

We devote the remaining few comments on Galilean invariance.  
Notice that by making explicit the dependence of the action (\ref{FEBO})
on the phase of the order parameter, the Galilean invariance of the action
becomes transparent. A Galilean transformation is described by the following 
change of variables:
$$
\begin{array}{clcrclcr}
\vec{r}'= & \vec{r}-\vec{v}t  &   &   & {d \over dt'}= &
\partial_{t}+\vec{v} \cdot \bigtriangledown \\
t'= & t  &   &   & \bigtriangledown'= & \bigtriangledown  
\end{array}
$$
where $\vec{v}$ is the velocity of the moving frame. 
Under a Galilean transformation
$\Delta(x)\Rightarrow \Delta(x)e^{-2i[m\vec{v} \cdot \vec{r}-mv^{2}t/2]}$
and therefore the phase changes into: 
$$\theta'(x') = \theta(x) - 2m\vec{v}\cdot \vec{r} + mv^{2}t $$
In the action (\ref{FEBO}) we can 
individuate two Galilean invariant terms:

\noindent i) $\left[ i\partial_{t}+i{\bigtriangledown\theta \over 2m}
\bigtriangledown \right]$, that can be interpreted as 
the "covariant" derivative,

\noindent ii) 
$\left[ \partial_{t}\theta(x)/2 + (\bigtriangledown \theta)^{2}/8m \right]$,
where $\partial_{t}\theta(x)/2 $ is the local chemical potential shift 
while $(\bigtriangledown \theta)^{2}/8m$ will give rise to the superfluid 
contribution to the kinetic energy of the system,
once we define the superfluid velocity 
$v_s=(\bigtriangledown \theta)/2m$.
 
Notice that, deriving the effective action (\ref{HUST}) we have spoiled 
the "covariant" derivative 
$\left[ i\partial_{t} + i\frac{\bigtriangledown \theta}{2m} 
\bigtriangledown \right]$ contained in 
the one-particle ``propagator'' (\ref{prop}).
As a consequence, for a given order in the perturbation theory, the Galilean
invariance for the effective action can 
only be recovered taking into account appropriate 
terms from higher order contributions \cite{Kemo,Aitc}.

\null


\begin{thebibliography}{999}
\bibitem{Tfin} We shall work in the real time framework of Ref.(7)
for a more
direct comparison with Ref.(12) even though the imaginary time formalism would
have been more transparent in dealing with the functional integrals.
\bibitem{NOT1} It is indeed known that the propagators of the HS fields are 
different from the propagators of the physical ones. As an example, the 
density-density propagator differs from the analogous auxiliary field 
propagator by a constant contribution (apart from a multiplicative factor 
that could be adsorbed in the definition 
of $\rho^{_{HS}}$): $$\langle \rho(x) \rho(y) \rangle = 
\frac{4}{U^2}\langle \rho^{_{HS}}(x) \rho^{_{HS}}(y)\rangle 
-\frac{2}{U}\delta(x-y).$$
\bibitem{over} Indeed, the choice $\alpha$=$\gamma$=1 gives the standard mean
               field results as saddle point solutions and, at the same time, 
	       guarantees that 
	       the Gaussian action reproduces the RPA approximation.
\bibitem{NOGA} Notice however that the lack of Galilean invariance of the mean
field equations derived for instance from the time dependent Landau-Ginzburg
theory is not an intrinsic problem of the time dependent Landau-Ginzburg
theory but of the mean-field approximation in the hydrodynamic regime.
\bibitem{NotV} The introduction of a lattice is the simplest way to introduce
               an ``ultraviolet'' cut-off to regularize the strong coupling 
	       limit.
\end{thebibliography}
\end{document}